# "Only the initiates will have the secrets revealed": computational chemists and the openness of scientific software

Alexandre Hocquet & Frédéric Wieber
LHSP – Archives Poincaré
UMR 7117 CNRS & Université de Lorraine
91 avenue de la Libération
B.P. 454
54001 NANCY CEDEX
FRANCE

## Abstract

Computational chemistry is a scientific field within which the computer is a pivotal element. This scientific community emerged in the eighties and was involved with two major industries: the computer manufacturers and the pharmaceutical industry, the latter becoming a potential market for the former through molecular modeling software packages. We aim to address the difficult relationships between scientific modeling methods and the software implementing these methods throughout the nineties. Developing, using, licensing and distributing software leads to multiple tensions among the actors in intertwined academic and industrial contexts. The Computational Chemistry mailing List (CCL), created in 1991, constitutes a valuable corpus for revealing the tensions associated with software within the community. We analyze in detail two flame wars which exemplify these tensions. We conclude that models and software must be addressed together. Interrelations between both imply that openness in computational science is complex.

-----------------------------

The quotation in the title, taken from a scientific mailing list, the "Computational Chemistry List", is intended to illustrate the tensions around the use of software in a scientific community. "Computational chemists", gathered around the uses of computers in chemistry [1], belong to a scientific field which started to grow in the eighties, whose aim was to develop "computational tools and techniques [which] offer a new method of attack in the continuing effort [in the chemical community] to obtain chemical information" [2]. Thus, the computer is a pivotal element of this scientific community, even though it is considered here as a tool, not as the object of the science in question. The adjective "computational", a typical word from the eighties and nineties, is essential: it is a scientific world of "computational science", not "computer science".

Numerous studies deal with the relations between computing and scientific activity, some of which are even considered as classics. Themes such as the "computerization of science" [3] or, for example, the mutual shaping of computing and biology [4] or the emergence of computerized evidence-based medicine [5] explore their interplay. Computational science has been addressed by scholars, for example the philosophical significance of its rise for scientific method [6], or the emergence of Monte Carlo simulations [7]. Many works also exist in the history of software [8], either on the software viewed as an industrial [9] or professional [10] activity, or on the difficulty and complexity of writing such a history [11]. Yet, software *per se* in computational science has attracted less attention, even if Spencer has conducted an ethnographic research within a computational fluid dynamics laboratory on a piece of software [12].

Our way to address the issue of software in computational science is to focus on application software in computational chemistry designed to model physico-chemical properties [13], a kind of software designed within the community and for the community. The actors are computer users in the sense that they do not create novel computing hardware or languages. They rely on the evolution of hardware and operating software designed by others. Yet, some of these "users" do code, some are software developers, sellers or even marketers. This community has a variety of different, mixed profiles when it comes to using or developing software.

The activity of developing computational tools in a particular scientific community leads to multiple tensions among the scientists involved as well in the development, the distribution and the maintenance as in the use of software. Developing, using and distributing software leads these actors to reflect on many things among which what scientific activity should or may be, what kind of relationship there is between scientific methods and the software implementing these methods, how their coding work is rewarded, which form of intellectual property to resort to, and whether to commercialize their research products. They also wonder about their ideal concept of the openness of science.

To make explicit these tensions, we rely on a corpus which is fit for revealing such tensions, namely a mailing list. This type of corpus has already been the subject of various studies in sociology and communication sciences [14], but very few in the field of the history of science. The informality of this kind of natively digital corpus allows unveiling the tensions between the actors, unlike the corpora of published scientific papers. In the context of this article, we present the so-called "Computational Chemistry List" that we use as a corpus, and we focus specifically on two extracted threaded conversations which show how the issues of methods, code, reproducibility of results, intellectual property and software marketing are articulated.

In order to be able to understand these tensions in a broader context, we first discuss the emergence of computational chemistry and its relationship to the computer. It is also important to understand the broader context of relationships between computational chemistry and the industry (the pharmaceutical industry and the hardware manufacturing industry) in times of mobilization of American universities to produce innovation [15].

These elements of context are fundamental in order to understand the specificities of the tensions raised by software in the particular scientific field computational chemistry is. Issues regarding openness in science, like reproducibility or epistemic transparency of methods, are specific in a scientific discipline involved in modeling, where calculability is a fundamental question. In our case study, the issue of parameterization of models is crucial knowing the particular history of theoretical approaches in chemistry. This issue has consequences for the topics of openness, reproducibility and epistemic transparency and leads to controversial situations among the actors as regards the interrelations between models and software. In this regard, the question of the openness of a science within which software packages are being developed, used and distributed is discussed in a particular way in computational chemistry. This field permits to understand these issues both epistemologically and socially. Moreover, our case study takes place in a specific technical, political and economical context that exacerbates tensions regarding theses issues: the democratization of the computer, the encouraged scientific entrepreneurship in Academia and the position of the scientific discipline between two powerful industries participate in tensions between academic and business norms, and between diffusion and robustness of methods. For these reasons, we believe computational chemistry is an interesting case study for addressing the interrelations between models and software: its context is specific enough to unveil the tensions posed by these issues, yet it allows to draw general conclusions regarding those interrelations.

After having set some elements of context, and analyzed the two threaded conversations we have chosen, we will then discuss the tensions at stake in order to conclude, finally, on our views about the roots of these tensions. We will argue that: 1/ models and software must be addressed together. Interrelations between both lead to the idea that transparency and validity of computational methods are complex, and that they are a source of tensions for chemists; 2/ The materialization of models into software is a way of spreading modeling methods in the broader field of chemistry (and not only computational chemistry), but a problematic way; 3/ Translating models into software in the broader context of relationships between computational chemistry and the industry leads to tensions between academic norms and software distribution norms.

## **Historical perspective on computational chemistry**

In order to understand the issues at stake in computational chemistry, we first dig into its epistemic roots. Scientists construct models in ways that are linked to the phenomena their models try to represent, to the theories they can use, but also to the technological, professional, economic and political context in which they work. These epistemic roots are a pivotal step to understand how the models they construct are then translated into software, and how these models influence the actors' discourses and the tensions at stake between them. In Mahoney's words, these models, and their translations in software, are "operative representations" [11] which are central to our study.

Computational chemistry has roots in the history of chemistry in at least two ways. On the one hand, "quantum chemistry" as a scientific field is the legacy in chemistry of the scientific breakthrough of quantum mechanics in the 1920s and 1930s. On the other hand, "molecular mechanics" as a field emerged in the 1960s in the sector of physical organic chemistry [16] and biophysics, in the era of the revolution of physical instrumentation (infrared spectroscopy,

NMR,...) in the chem lab [17].

The mathematical modeling of molecules is an idea which came up to chemists long before computers were available. Quantum chemistry is a scientific field that has existed since the first papers about the Schrödinger equation in the late 1920s. Theoretical physicists left quantum chemists with a very practical problem: to imagine theories (and models) to describe the molecules in a way that could be calculable and useful to the chemists [18]. They were the heirs of a reductionist worldview of microphysics, and the naming of the most popular quantum chemical theory ("ab initio") reflects the idea that the self-acclaimed scientific robustness of a model is based on the universality of its theory and tools, and should not have to deal with the tinkering of parameters, (or at least in their narratives) [19]. In practice, calculating was done with pencil and paper, desk calculators, and then with the use of excess computer time on the first supercomputers during the 1950s [20], with little rewards in terms of how big the molecules that could be actually calculated were [21].

Parallel to this, in the 1950s and 1960s, and because the computing facilities were attracting the interest of many scientific fields, a new kind of molecular structure theory arose, based on far simpler theoretical grounds, on a traditional conception of molecules in chemistry, and developed entirely on the pragmatic idea of tackling the modeling of what is actually computable. Organic chemists but also biophysicists showed interest in a theory based on a Newtonian classical mechanics view of molecules introduced by Infrared spectroscopists (a growingly popular method in the organic laboratory at that time [22]).

The benefit from this simplistic theory was the perspective to compute the properties of molecules ranging from the smallest to the most frequently encountered in organic, biological and pharmaceutical chemistry, by the computational standards of the time [23]. It was an ad-hoc modeling, based on tinkering parameters to fit experimental results, but an efficient one, to the detriment of the universality of the model: this ad-hoc modeling proved successful for a limited (but meaningful) number of molecular families (cycloalkanes, peptides, sugars,...) and the necessary parameterization to achieve results was at the expense of specialization. Each scientific team developed and parameterized their own method (a so-called "empirical force field"). Each team relied on different (and often competitive) protocols, based on different (and sometimes incompatible) spectroscopical or thermodynamical results, to actually define their parameters.

We can describe the situation as the parallel development of two different ways to model molecules. A first one is "quantum chemistry", concerned with the universality of their modeling, and impaired by the complexity of the mathematical and numerical formalism. The other is "molecular mechanics", concerned with the efficiency of computation to actually calculate some properties of significant molecules, and impaired by the sisyphal task of parameterization, and the fragmentation of methods.

Yet, the demarcation between them became blurred throughout the evolution of their respective fields, especially as the promises of the computer blossomed. During the 1960s, the so-called "semi-empirical" methods emerged: they were based on quantum calculations and thus formed a part of quantum chemistry, but they shared the idea of feasibility with molecular mechanics (the so-called "empirical" methods). In order to be actually computable, the quantum methods should be simplified, and above all, parameterized to achieve computability (the most lengthy calculations of the model should be replaced by empirical parameters). Similarly to molecular mechanics, different and sometimes competitive semi-empirical methods, based on different parameterizations, appeared in the 1970s.

In the words of Ann Johnson [24], these ways of modeling represent distinct "technical knowledge communities", from different disciplinary backgrounds, with different underlying theories, and even different epistemic traditions. Quantum chemistry as a keyword mentioned in publications, though rapidly growing through the 80s and the 90s, is superseded by molecular mechanics by the turn of the 90s. Molecular Mechanics, by the wider spectrum of studied molecules, became the favored modeling activity on the industrial side at that time and thus the first to give birth to a commercial activity of selling and licensing software in the field [25].Yet, all these ways of modeling were united by their common tool, the computer, and above all, they shared a common concern for *parameterization* of their models.

In 1974, The COMP "Computers in chemistry" division was created at the American Chemical Society (ACS). During the 80s, a new scientific field named "computational chemistry" emerged in the keywords used in scientific papers, in the conference calls for papers, but also in the academic and industrial research job offers or in the reports from supercomputing centers, "leading to the recognition of a new kind of chemist, different from a theoretical chemist, different from a physical chemist, an organic chemist, a spectroscopist or a biophysicist" [2].

## Computational chemistry and computers

The computer as a scientific tool had also evolved in the meantime, changing from an instrument of "Big science", a federally funded facility, one that was difficult to access both financially and technically, to a common device, one that would fit in every lab for many purposes: a networked, ready-to-use scientific tool which could be locally programmed and tinkered. Quantum chemists who had envisioned the use of computers to help in their calculations needed (and had) an inclination towards computing, but they also needed relations within the military bureaucracy and the computer science to persuade them to take advantage of the excess computer time on the supercomputers to use it for their own needs, something that none of the policy makers at that time could foresee as a promising computational application [20]. During this period of the 50s and 60s, the military and other government agencies funded computing science through grants and contracts, but also indirectly by buying the products of the computer manufacturers [26], and theoretical chemists were only a minuscule part of that plan. Yet, by the beginning of the 80s, computational chemistry had turned from a minor user of supercomputing facilities in the era of "federally funded Big Science" into a major client of the computing resources in the United States: 30% of the NSF supercomputing centers was dedicated to calculations in computational chemistry [20].

In the meantime, the transformations of the computer hardware were accompanied by transformations of the computer software. It was not until the late 60s that software became a product which could be purchased separately from the computer. Software development schedules slipped, costs rose tremendously, errors were harder to locate and correct with the increased complexity of codes, and revisions of the software were harder to implement [10]. Whereas hardware development was growing faster and faster, software development appeared to increase very slowly.

The development of the Personal Computer paved the way for a software industry, now that hardware was becoming standardized, thus lessening the problem of portability. In the 80s, most of the actions and profits were in the software business, which led to a shakeout of the industry into a few major players [27]. IBM was the dominant company in the computer industry in the 60s and 70s, whereas the software company Microsoft became dominant in the 90s. Through mergers and acquisitions, the software industry of the Personal Computer concentrated into a few dominant players (like Microsoft) with lock-in strategies to turn the users captive: a "winner takes all" market, very different from the mainframe software industry of the 70s.

This paradigm of the software industry in the 80s and the 90s was the computing world in which computational chemists were evolving, unlike the computer scientists that designed the Internet [28] and expressed strong opinions about software licensing [29] in a soon-to-be political movement of the Free or Open Source Software. It was also a world where software was becoming a dominant industry, an exemplary business adventure, including in the academic world.

In the era of workstations and personal computers, computational chemists merged computational methods into software packages. The aforementioned "technical knowledge communities" were then united by the computer as their tool, even more in times when it was becoming more accessible.

## **Computational chemistry and the industry**

The changes in the academic world in the United States at the beginning of the 80s have been described by many as a radical turn in structure [30] [31]. Especially relevant to our concerns is the change in the patterns of research funding, from a dominant federally (and above all military) funding to a "R&D competitiveness coalition" [32] that supposedly turned public scientific investigation into a pragmatic, profit-oriented activity, led by the idea that "innovation drives economy" [15]. It is well documented that, in large parts, the very structure of the University changed, with the creation of "patent or technology transfer offices" designed to make research activity, commerce, and innovation more and more compatible.

The eagerness to patent everything in the American universities had not waited for the Bayh-Dole act. Some Universities invested early in policies of technology transfer [15], but the influence of the Bayh-Dole act as a mechanics of change is relevant as the entire academic world had to follow the precursors' steps especially in the field of licensing or patenting academic software [33].

Computational chemists in Academia were thus stimulated or pressured by these transformations. They were involved with the industry, and this involvement was threefold: they were of course concerned with software as a business, because the software industry at that time was acting as a paragon of supposedly successful business models, and also because of the academic atmosphere leading scientists into entrepreneurship science. Computational chemists were also involved with the computer manufacturing industry. The giants like IBM were recognizing computational chemistry as a new major force in the field of supercomputing. For example, IBM was a major investor in one of the big molecular modeling software vendors, Polygen [34]. Computational chemists were also one of the best customers, (and advertisers) for those who manufactured workstations with high graphical capabilities like Silicon Graphics. The third reason why corporations with academic computational chemistry background were created during the 80s is that a potential market for computational chemistry modeling had been envisioned by the pharmaceutical chemists and the hardware vendors. Structure calculations were viewed by pharmaceutical industry R&D departments as a "technological promise" [35] of potential savings in the discovery and assessment of new drugs (Rational Drug Design) in a context of ever rising costs of new drug leads [36]. Corporate computational chemistry teams dedicated to pharmaceutical research were created in-house, or corporate funding was invested into academic groups to produce results or to develop software.

The influence of the pharmaceutical industry on computational chemistry as a scientific field was also a cultural one: the most successful narratives of entrepreneurship science came from the neighbored field of biotechnology [37], a scientific domain itself involved with the same industry. Furthermore, the pharmaceutical scientific field and industry share a culture of secrecy and patenting rather than publishing and sharing results, which consequently influence the scientific fields which deal with them [38].

Software developed by computational chemists was turning from "user-oriented" programs to "market-oriented" packages. The merging of methods into packages were conceived with the aim to enlarge the user base. Software was once programmed and then sent to the QCPE (Quantum Chemistry Program Exchange) to be given away to any interested party for free and "as is" [39]. It was now planned, designed and developed to be distributed in the academic and industrial disciplines of chemistry that could benefit from computational methods. In the eighties, the number of publications using commercial computational chemistry software grew exponentially, as grew the number of chemistry calculations published in the industry [25].

## **The Computational Chemistry List (CCL)**

It is in this context that the Computational Chemistry (mailing) List (CCL) was created in January 1991 by Jan Labanowski, a computational chemist, then an employee of the Ohio Supercomputing Center (OSC). The purpose of the list was to gather a fledgling community of researchers. As computational chemistry was a field in its infancy, the chemists willing to use computational tools lacked education in the field, and the scientists who developed these tools found a unique way to disseminate them. The primary goal of the CCL was to "educate and get educated" [40]. The rules of this list, as defined by the moderator, allow anyone to contribute, making the CCL particularly inclusive. From graduate students to senior researchers, from code developers to « end users », from hardware vendors to software marketing sales forces, the CCL was (and still is) the arena where all the people linked one way or another to the molecular modeling software could debate. It is hard to assess the representativity of the population of the CCL subscribers in terms of social or professional profiles within the computational chemistry community, but it is safe to say that grossly each profile has a loquacious enough character among the CCL subscribers to speak up.

The CCL grew steadily in terms of number of subscribers and number of daily messages from 1991 to 1995, when it reached a plateau of several thousands of subscribers and a dozen daily messages. The topics encountered in the CCL, apart from announcements of academic events and requests for literature, are opinions or help wanted on scientific topics [41]. The main kind of topic is a request for help in using software. The CCL was more often than not the quickest way to find help from peers, creating an atmosphere of mutual aid but also sometimes an atmosphere of resent when commercial software is accused of outsourcing its maintenance duties. Yet, commercial announcements are explicitly allowed in the CCL unlike in most academic forums: this epitomizes the intimate relationship between the academic and economic worlds in the field of computational chemistry.

It would be naive to view the CCL as a public sphere with abolished hierarchies. A few anthropological studies of mailing lists have shown how issues of gender or more generally issues of differences of status can interfere in, or even structure an Internet based conversation [42]. It is untrue that no relationships of power or authority exist within the participants of the list, but it is also true that they are different from what they are within a laboratory, or in a conference, or in the process of publishing a paper, and they have thus led to new forms of interactions in the debates. As Grier and Campbell put it in their study of Listserv [43], the mailing list is a place where the participants of the list interact within the community without acting in front of an audience.

Topics which evolve into passionate debates, or even heated arguments, can be valuable pieces of information from a historical perspective, even though they are unhealthy for the list itself. Given that discussion is possible and even encouraged if not considered off-topic, then the most controversial tensions within the community generate interesting threaded conversations where a variety of actors within the community can interact. In a similar way that "scientific controversies" are interesting for STS scholars to learn about the scientific, political and social matters at stake, "flame wars" (the threaded conversations in which the topic is controversial enough to degenerate into a self-sustaining avalanche of posts [44]) represent a way of identifying which topics are actually a source of tensions within the community. Provocative posters, or "trolls", tend to disrupt the harmony of the community by posting on controversial topics, thus forcing the community to react, degenerating into a flame war. Yet, the flame war incites the community to discuss and debate about sensitive matters, forcing the members out of a polite stance and thus leading them to reveal otherwise concealed opinions [45].

In this regard, the threaded conversations provide an interesting material for investigating the actors' day-to-day practices and discourses in a kind of microhistory. In this study, we focus on a qualitative analysis of two chosen heated threads which illustrate the issues as regards software and their evolution throughout the 90s. The comparative analysis of the debates and arguments, of the context and the actors, provides substantial information to characterize these issues.

## 1993: the first CCL flame war ever

The first conversation thread we want to discuss in order to disclose some of the tensions produced by software within the chemists' community starts on 06/23/1993, with a seemingly innocuous message. Twenty-nine posts from eighteen subscribers will follow for ten days. This thread constitutes the first flame war ever on the CCL. In comparison with contemporary flame wars, the number and density of the messages is relatively small. 1993 was an era of bandwidth frugality. The first message is an announcement. Andy Holder, then Assistant Professor of Computational/Organic Chemistry at the University of Missouri-Kansas City and president of a scientific software company named Semichem, Inc., announces the publication of a paper providing results for a new quantum chemistry semi-empirical method named "SAM1". In this message, Holder writes down: "This [the paper] is primarily a listing of results for the new method for a vast array of systems. [...] A more complete paper describing the model will be forthcoming" [46]. This is the last sentence quoted here which will launch the debate. Graham Hurst (then working for the software company Hypercube, Inc.) wrote in the second message of the thread: "this [Holder's] post disturbs me..." [47]. Hurst considers that "it will be impossible to independently reproduce these results" because the model leading to the results has not already been published. He adds: "If the method has not yet been published, then the results should not have been accepted for publication since they cannot be verified". Thus, the initial problem of the flame war is an epistemological problem associated with a problem of publication ethics: as the details of the model used to produce the published results have not been published, the results cannot be independently reproduced and verified and are then not considered as publishable.

Figure 1 portrays the dynamical organization and ramifications of the thread, showing who is responding to whom. Three directions of discussion are opened up in response to Hurst's post. First, the question of how possible it is to verify the validity of the results is discussed as the possibility to reprogram the computational method by oneself. Is the information necessary to reprogram the method available? A discussion then opens up regarding more generally the issue of the parameters that are used in semi-empirical methods. These parameters are central in the different semi-empirical methods used and they are not always made publicly available. They are sometimes hidden away in the source code of the program, which is not always made public. As one of the participant of the thread writes down: "[...] we should like to know your opinion on the actual trend in commercializing computational packages without source codes. Does this trend encourage the development of science? And also: up to what limit a computational package can be considered as a product of a single research group?" [48]. Thus, the epistemological question of verifying the results is associated with the questions of the openness of the source code, of the commercialization of computational packages, and of the computational methods as a scientific public good (a public good at the disposal of the community, but equally produced by it).

In the second direction of the discussion, the tension between the world of academic research and the world of scientific software corporations is underlined. In response to Hurst's message, Holder concedes that it is not always easy to clearly distinguish scientific from entrepreneurial activities. The scientists' implication in scientific software corporations, along with the costs necessary to develop software, telescopes the values (openness, reproducibility) the actors associate with science. As Holder puts it: "So, while Dr. Hurst's point is well-taken and fully subscribed to by me both in my capacity as a university researcher and president of Semichem, there is no intention to "hide" anything. I understand the sensitivity of this issue and I am committed to the pursuit of science in an *open atmosphere*. [...] The development of SAM1 is my primary research activity at UMKC, but Semichem is also spending money to develop this method and will be giving it to the scientific community freely. We withhold only our code. [...] It should be noted, however, that *some interests are not scientific, but competitive*" (emphases added) [49].

In the third direction of discussion, the problem of publication ethics is discussed. The importance of the peer review process in scientific publishing is underlined and some contributors ask if reviewers do a good job when accepting for publication results which have been obtained by a computational method not fully (and openly) described. The question leads more generally to contrast proprietary methods and open scientific literature. As Mark Thompson, then research scientist at the Pacific Northwest Laboratory and developer of a freely licensed molecular modeling program called Argus, writes down: "I feel very strongly that when a new method is developed and implemented that it must pass the peer review process to gain legitimacy in the scientific community, regardless of whether most other scientists care to re-implement that method or not. Proprietary methods are fine, as long as it is openly known that they are proprietary. Results of proprietary methods do not belong in the open scientific literature" [50]. Of course, these three directions of discussion are interrelated.

The sixth message of the thread, written by Douglas Smith (then Assistant Professor of Chemistry at the University of Toledo) is particularly revealing. In this long post, Smith responds point-by-point, using interleaved posting, to Thompson's whole message. The tensions produced by software within the community are interestingly expressed by contrasting how scientists believe they should act with what they actually do. Thompson has written that "good science is that of reproducibility and independent verification" [50]. Smith points out that it is "universally true and accepted" but "rarely followed" [51]. Smith uses as an example the issue of parameters used in molecular mechanics, which are regularly modified and adjusted for a particular study without being published in the paper relating to that particular study. More generally, the very nature of such a method (and of semi-empirical methods) leads to a multiplication of the parameters used without a clear display of which parameters are used when producing such or such results. The problem is then more general than for the single case of the "SAM1" method. In practice, chemists act in a way that differs from what they say they should do. Thompson has also written: "If the results of a new method are published without sufficiently describing the method to fulfill the above criteria [reproducibility and independent verification], then I personally could not take the results seriously" [50]. Here again, Smith considers that if this position points to "a real problem", it is "utopian and most likely not practical", because of "the proprietary nature of commercial software" [51], and because some people use this type of software as a "black box". He then adds: "Besides, who ever said we had to reveal all our secrets and make them readily available and accessible? When software copyrights and patents really provide adequate protection, maybe I will agree with that attitude" [51]. Finally, if "results of proprietary methods do not belong in the open scientific literature", as Thompson has written, "where do they belong?" Smith replies. According to him, the situation is complicated: "what about the difference between someone in industry who paid for the source code for MacroModel as compared to the academic, such as myself, who only gets binaries? Are my results to be less acceptable because I don't have the absolute method available? Or are the industrial results less acceptable because they can be the results of tweaking the code?" [51].

In Smith's post, the discrepancy between the kind of values (openness, reproducibility) the actors associate with science and their actual practices associated with computational methods and software is clearly highlighted. Because of the very nature of (semi)empirical methods, which lack epistemic transparency, because of the proprietary nature of some software packages, because of the possibility to use software as black box, the question of the norms of sound science is in practice difficult to resolve. Moreover, computational chemists ask the question of how the difficult and tedious work of programming can be recognized. Can this recognition be obtained by publishing programs or by adequately protecting them ("When software copyrights and patents really provide adequate protection […]" Smith writes)? The complexity of the issue of software copyrights and patents is then stressed in many subsequent posts of the thread. The mentioning of patents, copyrights and licenses in numerous later posts is often done on an interrogative mode, and the thread finally dies of attrition after a general sense of uncertainty about what the future holds regarding the relationships between these intellectual property notions and the tensions they expressed beforehand.

## **2001: the great Gaussian flame war**

The second conversation thread we want to discuss starts on 12/05/2001. Forty-five posts from thirty-three subscribers will be sent in the following seven days. In comparison with the first thread, the number and frequency of messages is higher, reflecting the change in email usage. Here again, the first message is an announcement which seems to be innocuous. Jen-Shiang Kenny Yu, then Ph.D. student in the Department of Chemistry of National Tsing Hua University (Taiwan), indicates in his message that the results of a benchmark performed in his lab, for PC computers, of the "popular electronic structure program Gaussian" are made publicly available on a webpage. This benchmark has been carried out for several combinations of microprocessors and random-access memory devices [52]. If computing in computational chemistry was mainly performed on workstations in the 1990s, this benchmark shows that desktop computing is going to break through in the 2000s.

Figure 2 portrays the dynamical organization of the thread, just like Figure 1. The messages in the 1993 thread were mainly pluritopical, the different branches of the thread being interrelated, whereas the messages of the 2001 thread are mostly monotopical. After his first announcement message, Yu posts three other messages. His second and third messages show that the benchmark, and notably one technical question, interests many people. As Yu writes down in his third message: "There are several persons asking about the makefile [53] to compile Gaussian 98 with Intel Fortran compiler" [54]. However, Yu also indicates, in this same post: "We'll post the detail [of the makefile] on our website after we make sure that it won't violate the license agreement of Gaussian" [54]. This is this question of violating the license agreement that will set the thread on fire. In his fourth and last message, Yu thus writes: "We have got the information from Gaussian Inc. that distributing the modified version of makefile or the instructions is violation to the license agreement" [55]. From this starting point, the discussion is launched on what the license of scientific software can or must allow, and the flame war begins with the intervention of the CEO of Gaussian, Inc., himself, Mike Frisch.

The first reaction to Yu's last message comes from Richard Walsh, then Project Manager in Cluster Computing, Computational Chemistry and Finance for netASPx, Inc.. He writes: "What about a simple description of how to do it without any lines directly copied from the file? That is your intellectual property which I assume that you are free to distribute?" [56]. Then, the policy of Gaussian, Inc. is challenged, in a deliberately provocative way, by Chris Klein (then at the Department of Applied Biosciences, Pharmaceutical Chemistry, Swiss Federal Institute of Technology / ETH Zurich), by metaphorically translating Gaussian policy to the automotive business: "[...] the company's policy, translated to the automobile business, appears to be: "OK, we'll sell you the car (program), but you have to produce the proper key (makefile) yourself.. if you copy the key from someone, we'll sue you... maybe we can give you the key for the trunk. Rather strange way of doing business" [57]. In a third reaction to Yu's last message, Serguei Patchkovskii (then Research Council Officer in the Theory and Computation Group of the Steacie Institute for Molecular Sciences, National Research Council Canada, Ottawa) argues that Gaussian license is very restrictive: "Taken literally, this license prevents you from even -posting- Gaussian output to this list (or from providing it as a supplementary information in a scientific publication), for two reasons: a) it discloses performance data, and b) the output may prove to be of use to one of Gaussian, Inc. competitors" [58]. The issue of the licensing policy of Gaussian, Inc. is thereafter pivotal in the remaining of the thread.

Mike Frisch replies to these criticisms, and notably to Klein's provocative automobile analogy, by pointing out that, unlike many other software vendors, Gaussian, Inc. provides the source code of Gaussian as well as "[...] makefiles for supported platforms and compilers" [59]. He then adds that making the program run on other platforms, with other compilers and makefiles which have not been tested by the company will lead, if made public, to unreliable versions of the program being used and then to problems for the technical support of Gaussian. He concludes: "The normal way of doing business, which is what most of our competitors do, would be to not license the source code at all and hence not be subject to criticism of the terms of the source license" [59]. His defense is then articulated around the availability of the source code, which is seen as being fundamental for scientific software because it allows epistemic transparency, and around the question of the support Gaussian, Inc. has to offer to its purchasers and users.

The thread then splits into two topics: 1/ the issue of the technical support and user-friendliness of Gaussian; 2/ the articulation between the availability of the source code, the possibility (or not) to implement the software on different platforms, and the stability of the software associated with its protection by Gaussian, Inc.. Regarding technical support and user-friendliness, the discussion starts with a post by Max Valdez (Maximiliano Valdez González, National Autonomous University of Mexico). He writes down: "I think Gaussian is a great tool, but it has a LOT of little secrets and "bugs", and the support is not so good […]" [60]. He tries to refute Frisch's argument about the problems that will emerge for the technical support of Gaussian if unreliable versions of the program were used. He adds that most of the questions he has asked to Gaussian technical support have finally been answered when asked on the CCL list. If a community of users constitutes a more effective support than the official support, then why "Gaussian doesn't have *a more open policy* to allow end users to communicate improvements, or new ideas specially for new compilers and boxes" (emphasis added) [60]?

This discussion about support and end users then leads to the question of the user-friendliness of Gaussian. Phil Hultin, then Associate Professor of Chemistry at the University of Manitoba, Winnipeg, indicates that he represents "a new kind of scientist working with tools like Gaussian" [61]. Hultin is an experimentalist (in organic chemistry), not a "computer whiz" (his expression). And for end users such as him, Hultin considers that an effort has to be made in order to improve the quality of the Gaussian manual and to provide a user-friendly interface. Hutlin's post shows that computational tools and software are being democratized at this time in chemistry, in particular because they can be implemented on computers which are cheaper and more accessible. Several messages then discuss this democratization phenomenon. Hultin's suggestion is for example criticized because it would lead to so-called "black-box" software. Some computational chemists ("computer whizzes" in Hultin's words) reject such black-boxes because they lack epistemic transparency. We can see, here, that "open" in the sense of providing epistemic transparency by making the source code readable is different from "open" as empowering larger audiences of end users by providing more transparent software for such users. Even if he later concentrates on the issue of the user-friendliness of Gaussian, this is this entire complexity of the issue of the openness and transparency of Gaussian that Hultin tries to express when he writes "maybe people would be less prone to jump on Gaussian (as they have over this makefile thing) if they didn't feel that the philosophy of Gaussian Inc. was similar to that of the "high priest" - only the initiates will have the secrets revealed and then only after years of study" [61]. The topic of the second branch of the conversation thread shows another dimension of this complexity. "Open", in the sense of providing epistemic transparency by rendering the source code readable, is not satisfying for all "computer whizzes". The discussion continues with messages asking why Gaussian could not take into consideration the makefile Yu has developed, and try to test it. The debated question is then to know who can contribute to Gaussian, and how it can be used. Is scientific software a public good, which has to be developed, maintained and enhanced collectively, or is it a private product whose protection against derivatives guarantees its stability? Finally, the discussion ends with posts which question the significance of having the source code available if it is not possible to use it on different systems by modifying the makefile [62], and posts which ask for easier compilation routines in order to increase the portability of the software, as well as the availability of a comprehensive test suite in order to verify the compilation, as a desired sound scientific practice [63].

We now want to focus on which issues are unveiled out from the debates within these two threads.

## **Multiple tensions**

The epistemological nature of the models in computational chemistry implies that epistemic transparency is an ideal vision of modeling. The very nature of the models, for example in semi-empirical methods like SAM1, requires a time consuming work of parameterization. Parameterization poses a problem of reproducibility and transparency. Scientific parameters designed to make the model actually produce robust results possess their own epistemic problems (like calibration, theory groundedness, fitting, (lack of) universality...). But there is more: parameters are also intertwined with the coding of the method to make the program run. The entanglement of scientific and coding parameters is turning the concept of reproducibility into a problematic issue because of the complexity of the code. The consequence is that the reproducibility as well as reprogramming of a method, even with an open source code, is highly unlikely, and even more so with a mere publication in hand. This *epistemic opacity* is a source of tensions and is criticized for example from the point of view of experimental chemistry. This epistemic situation, and the tensions implied by parameterization, are constitutive of computational chemistry.

The lack of transparency is also present in other aspects of software. Not only do software vendors sometimes choose to sell (or license) only executables/binaries, but when they do provide an open source code (to comply with an epistemically sound science), the accompanying licensing strategies may consist in prohibiting to manipulate/modify/reveal/benchmark/test said source code, generating frustration among end-users. In other words, code openness is criticized as worthless if it does not go along software openness: if other scientists cannot compile, test, benchmark the code, then the transparency issue is also an interoperability issue. Software is thus also linked to hardware.

A tension exists between the desire for computing power and efficiency (in a rapidly evolving hardware world) and the epistemic robustness linked to the scientific software: it must produce sound results in a growing variety of hardware conditions. This tension arises in times when hardware becomes increasingly available in the laboratory. In 2001, the Personal Computer became a common scientific tool in the laboratory, and also a tool that one can tinker to improve in-house performance. The great Pentium vs AMD processor competition (typical of those times) translates into desire for benchmarking the modeling software with diverse hardware environments, something that restrictions to compilation hinder.

Tensions also proceed from the confrontation of an idealized scientific world with a scientific world in a context of software. Academic publishing, which constitutes the traditional form of academic reward, is central in the actors' ideal concept of the openness of science. Yet, as software is more than just code, but as well a commodity, the scientific activity then shares common concerns with an industrial sector. In a world where software is also a business, issues of intellectual property, or software distribution in general mix with the traditional concerns of the scientific world. The difficulty to finance continuing development or the difficulty to establish a serene relation with users/customers regarding development and maintenance choke with scientific ethos concerns. This leads, for example, to the question of the lack of scientific recognition for software development in the 1993 conversation thread, or to the discussion of the effectiveness of the technical support of Gaussian in the 2001 thread. In this regard, the developers (and vendors) who license their proprietary software with an open source code but draconian restrictions to its use are trying to limit a potential exploitation of their code by competitors, but they also argue that a compilation of the software that would not comply with the in-house rules could lead to unsound scientific results.

In the narratives of the posters, this issue is expressed as "science as a public good" versus "software as a commodity". The fact that software, as the materialization of a scientific model is developed "with taxpayers' fund" (a popular expression in the threads) by academics, appears to many as conflicting with software viewed as a business model. Whereas academic institutions may promote so-called technology transfer by fostering scientific entrepreneurship for computational chemistry software, the idea that the federally funded development of scientific models is turned into a business model (and a promising one for corporate molecular modeling, especially in the pharmaceutical industry) is frowned upon by a large part of the community, and this raised concerns particularly in 1993. There is a tension between the idea that modeling software, as a scientific tool, should be considered a public tool, and as such, one that belongs to the scientific community, including in its potentiality to be enhanced (and maintained), and the idea that, as a tool developed by a small team, in a commercial context, strict licensing policies help to keep software stable, which guarantees the production of sound scientific results.

Especially for "commercialized" software, the issues of (lack of) maintenance and support and the issue of the "black box" syndrome further divides the community in other terms: first appears the issue of different kinds of software users, then follows the issue of different kinds of support for corporate and academic users, and even the issue of different kinds of modeling parameters (and different levels of secrecy) for the same method in corporate or academic environment. There are also different kinds of users in terms of computing literacy, leading to a divide between lay users and "computer whizzes". The lay users express their concerns (and they often find it hard to achieve legitimacy on the list) from the viewpoint of an experimental chemist who wishes to use computational tools: they need to appropriate the software and they have a hard time doing so as they are easily accused of turning modeling into unsound science (by merely pushing buttons), and they find themselves delegitimized. The tension here for software developers lies in the dichotomy between augmenting their users/market shares and keeping the control on software (as a method, as a code, and as a commodity) [64].

Throughout the nineties, molecular modeling software shared issues with a growing industry of software in general. More precisely, the molecular modeling software, as a promising technology, was promoted by hardware vendors and bought by the pharmaceutical industry: the concern with the software "business model" was preeminent for the CCL posters. References to "the market" as a regulating force is not uncommon in CCL messages to characterize what a sound business practice should be. Yet, this scientific software targets scholars, some of them working in the (pharmaceutical) industry, some of them in Academia. It is a market niche, and one where potential customers have very different resources. In particular, the software support that the former can afford and the latter cannot, is a dividing topic. On the other hand, this scientific software is sometimes designed in a corporate environment, sometimes in an academic one, and sometimes in an academic structure which turned into entrepreneurship.

Finally, debating is phrased in 1993 into publication reward, acknowledgment of coding work, copyrighting and patenting concerns about software, and CCL posters express uncertainty regarding these issues. In 2001, tensions now arise with a technical argument over a particular software licensing policy which integrates a very large part of these issues, a business conflict between the software and its competitors, a conflict between the software and some of its users, a conflict between different categories of users, a conflict dividing the scientific community. It is striking that in 1993, questions are asked about methods, and reproducibility issues are expressed in abstract terms of publication and code openness. In 2001, the open source code issue was a very practical one and the tension arose from the inability to create makefiles due to licensing restrictive policies. Licensing then implies multiple concerns: scientific concerns (the source code of Gaussian is readable in order to preserve epistemic transparency), intellectual property concerns (rewarding the work of programming), and business concerns (restrictions to the use of the source code to limit its potential exploitation by competitors).

**Roots of tensions**

From the discussion of these multiple tensions, we now dig into their roots. Our first point is an epistemological one: in a computational science, models and software must be addressed together. Interrelations between both lead to the idea that transparency and validity of computational methods are complex, and that they are a source of tensions for chemists.

If it is interesting and necessary to discuss the structure, properties and epistemological status of models, as it is common in the philosophy of science, we think that it is necessary to understand models in relation with software which embody them, which give them their productivity. In turn, understanding software (in computational sciences) needs to take into account the models they express, that is "the representations of world" scientists translate in a way the computer can "understand". These representations depend on the communities of scientists involved and the histories of the ways they represent the portion of the world they are interested in [11].

The interest of discussing both models and software can bee seen in the specific relationships between computational chemistry models and computational chemistry software. The complexity of the parameterization is central in the modeling activity. This fact has to be understood in the context of the calculability problems quantum approaches in chemistry have faced. In the case of molecular mechanics methods, the choice of a particular representation of matter, which is consistent with a classical conception of molecules, also leads to a necessary complex work of parameterization. The choices of sets of parameters, made locally by such or such research group for such or such group of molecules, lead to models whose epistemic transparency is questioned by the actors themselves. What is interesting for our argument is that this lack of transparency of models has repercussion on the status of software: the question of the openness of the source code is for example made more salient in the 1993 flame wars knowing the importance of parameterization in modeling. Finally, materializing models into software could lead to black-box models into software as a scientific push-button instrument. Many actors fear this perspective, which deepens in return the question of the the lack of epistemic transparency of the models. This kind of bi-directional repercussions shows clearly that discussing both models and software is crucial.

As the 2001 conversation thread shows, this materialization of models into software also leads to address the validity of methods, not only in terms of model transparency, nor transparency of software as openness of the source code, but as well in terms of how the software is compiled (this implies hardware concerns). The epistemic validity of the scientific results produced when running software is thus not just an issue of translating models transparency into software, it is also entangled with compiling software, hence the issue of benchmarking software for various hardware configurations. The software licensing and maintenance policy is then in question. It could be seen as a protection, developers and maintainers being responsible of the reliability of the scientific tool the software is. But it could also be criticized as impeding the performance and reliability of scientific software.

Our second point is about discipline dynamics: a second kind of tensions arise between two groups of actors, well represented in the 2001 thread. The first group is constituted of the scientists who develop and/or use as expert-users computational chemistry software. The other one is represented by Phil Hultin, an experimental chemist and lay-user of computational tools. Tensions arise between this two groups because as the second group ask for more user-friendliness, the first group fear a phenomenon of using models as black-boxes. This fear is associated with the epistemic status of models, as already discussed. But it has also to be understood in the broader context of the somewhat difficult recognition of modeling and simulation as sound science in the whole field of chemistry. Computational chemists have to be particularly cautious concerning the question of the validity of the results they produce in order to gain credit. However, distributing more user-friendly software can be a way to enlarge the community of users and then to gain recognition in the chemists' community.

The materialization of models into software is in this manner a way of being recognized in chemistry, but a problematic way, and this leads to tensions. In this sense, the question of the recognition, trustworthiness and diffusion of computational chemical software in chemistry can probably be analyzed as the adoption of a new instrument, which has to be constituted as trustworthy but also user-friendly.

Our last point regards social norms: translating models into software in the broader context of relationships between computational chemistry and the industry leads to tensions between academic norms (publishing as reward) and software distribution norms (licensing, commodification).

In the 1993 thread, chemists typically associate transparency with open scientific literature, publication being classically viewed as the major form of reward in academic norms. As coding is not rewarded within academic norms, business norms of software commodification are used by some developers. This leads to the issue of licensing software and to the debated question of knowing if a scientific software is a public good, which has to be developed, maintained and enhanced collectively, or a private product, whose protection against derivatives guarantees its stability. The clash between these two types of norms is manifest in the 1993 thread when being expressed by contrasting how scientists view openness as an ideal scientific value and their actual practices. The relations with the pharmaceutical industry, with its culture of secrecy, exacerbate these tensions.

## **Openness**

Openness has been ubiquitous in our account of the threaded conversations in the CCL, and more generally in computational chemists' concerns over the years. It has been employed by the actors, but also in our account, with many different meanings. These meanings, and the ambiguities associated with this polysemy, are revealed by the tensions highlighted in our study.

Openness is to be understood first of all as an ideal value associated with an ideal vision of science. It is often referred to by the actors as an essential norm, but also as an argument to distinguish between ideal science and real practices. This first meaning is associated with a more practical meaning of openness as epistemic transparency and reproducibility of methods, associated with the concept of publishing in the scientific literature. The role of publication as scientific reward and of reviewing process as the warrant of the validity of scientific results is being discussed in this context.

Yet, the epistemic transparency of the scientific methods developed is blurred by the fact that these methods are entangled with their programming. Another meaning of openness is the openness of the source code, and our examples show that this is understood in two ways. First, associated with epistemic transparency, an open source code is a readable source code. But the 2001 thread shows that this widely accepted meaning is criticized as being not enough, the readability being considered by some actors as useless without the possibility of compiling/testing/benchmarking.

Moreover, the licensing policies that software developers choose to adhere to regulate the mutual shaping of methods, software and hardware. The next meaning of openness is thus about the openness of the policy of the software: how the licensing policy frames the practices of the end-users and how the corporate policy (especially regarding support and maintenance) shapes diverse categories of users.

At a sociological level, the issue of how inclusive the community of computational tools users is is debated. End-users want user-friendliness in order to become empowered whereas lead-users argue against computational tools as black-boxes in order to preserve scientific soundness. The issue of different levels of support/maintenance or even licensing underlines the various statuses of the actors: academic or industrial users, academic developers, corporate developers and vendors. The last meaning of openness is thus understood as empowerment of categories of users.

Finally, beyond the study of a particular computational scientific field, our story addresses the general issue of "scientific openness" as a blurred concept. "Openness" may have even different meanings in other contexts, and it is of course beyond our study to address them all. Yet, our study helps to highlight that addressing the complexity of openness in computational models requires to take software into account in its many aspects.

2, pp. 254–280, Apr. 2007.

**Figure captions**

Figure 1 portrays the 1993 thread structure. Each node represents a post (with the name of the author and the date). Each edge represents the citation of a previous post. Grayed out posts are the ones quoted in the text.

Figure 2 portrays the 2001 thread structure. Each node represents a post (with the name of the author and the date). Each edge represents the citation of a previous post. Grayed out posts are the ones quoted in the text.

-------------------------

**Short author bios**

Frédéric Wieber is maître de conférences in History and Philosophy of Science at the Université de Lorraine (Nancy, France). He is a member of the Laboratoire d'Histoire des Sciences et de Philosophie – Archives Henri Poincaré, UMR 7117 CNRS – Université de Lorraine. He has a PhD in History and Philosophy of Science from Université Paris Diderot. His works include papers on the history of computational protein chemistry in the 1970's and 80's and on the calibration of scientific instruments. He is more generally interested in the tools used in theoretical and computational scientific practices. Contact him at frederic.wieber@univ-lorraine.fr

Alexandre Hocquet is an ex computational chemist academic and now a Professeur des Universités in History of Science at the Université de Lorraine and a member of the same laboratory. His focus is on STS, particularly the relationships between software and production of knowledge with works on computational chemistry, but also Wikipedia and Football Manager. Methodologicallly, his works rely on the analysis of threaded conversations in webforums or mailing lists. More at http://poincare.univ-lorraine.fr/fr/membre-titulaire/alexandre-hocquet Contact him at alexandre.hocquet@univ-lorraine.fr Follow him at @osvaldopiazzoll